\documentclass[fleqn,12pt,twoside]{article}
\usepackage{espcrc1}

\usepackage{graphicx}
\usepackage{epsfig}
\newcommand{\AmS}{{\protect\the\textfont2
  A\kern-.1667em\lower.5ex\hbox{M}\kern-.125emS}}

\def\lsim{\raise0.3ex\hbox{$<$\kern-0.75em\raise-1.1ex\hbox{$\sim$}}}
\def\gsim{\raise0.3ex\hbox{$>$\kern-0.75em\raise-1.1ex\hbox{$\sim$}}}

\title{
Hadron correlators, spectral functions and thermal dilepton rates 
from lattice QCD\thanks{The work
has partly been supported by the DFG under grant FOR 339/2-1.}}
\author{Frithjof Karsch$^a$, 
S. Datta$^a$, E. Laermann$^a$, P. Petreczky$^a$, S. Stickan$^a$
and I. Wetzorke$^b$
\\
\vskip 6pt
$^a$ Fakult\"at f\"ur Physik, Universit\"at Bielefeld, D-33615 Bielefeld, 
Germany\\
$^b$ NIC/DESY Zeuthen, Platanenallee 6, D-15738 Zeuthen,
Germany
}      
\begin{document}

\maketitle

\begin{abstract}
We discuss 
information on thermal modifications of hadron properties which can
be extracted from the structure of Euclidean correlation functions of 
hadronic currents as well as more direct information obtained
through the reconstruction of the spectral functions 
based on the Maximum Entropy Method.
\end{abstract}

\section{Introduction}

Euclidean time correlation functions, $G_H(\tau, T)$, of operators with
hadronic quantum numbers $H$ carry all the information about the hadronic
spectrum in this quantum number channel, e.g. about the temperature dependence
of hadronic masses, the width of these states and eventually also about
their disappearance from
the spectrum which may happen at high temperature. This information is 
carried by the spectral function, $\sigma_H(\omega,T)$, which is related
to $G_H(\tau, T)$ through the integral equation,
\begin{equation}
G_H (\tau, T) = \int_0^{\infty} \hspace*{-0.2cm}{\rm d} \omega \;
\sigma_H(\omega,T)\;
{ {\rm cosh} (\omega(\tau -1/2T)) \over {\rm sinh} (\omega/2T)}
\; \equiv \; \int_0^{\infty} \hspace*{-0.2cm}{\rm d} \omega \;
\sigma_H(\omega,T)\; K(\omega, T) ~.
\label{correlator}
\end{equation} 
At $T=0$ the kernel, $K(\omega, T)$, reduces to an exponential function and at 
large
Euclidean times $\tau$ the correlation function picks up the contribution
from the lowest lying state contributing to $\sigma_H$, {\em i.e.}
$G_H(\tau, T=0) \sim \exp(-m_H\tau)$. At non zero
temperature many excited states contribute to the correlation
functions, the in-medium spectrum of hadrons gets modified through 
interactions among hadrons, and an analysis of the asymptotic behaviour
of $G_H(\tau, T)$ is no longer possible as $\tau$ is limited to the interval
$[0,\; 1/T)$. The reconstruction of $\sigma_H(\omega, T)$ and in 
particular the determination of its low energy structure thus is 
difficult at non-zero temperature. Additional complications arise
in lattice calculations which necessarily are performed on lattices 
with finite number of points ($N_\tau$) in Euclidean time. The correlation 
functions 
$G_H (\tau, T)$ can thus be calculated only at a finite set of Euclidean times, 
$\tau T\; = \;k/N_\tau$, with $k= 0,\;..\; N_\tau -1$. 
In order to reconstruct the spectral
functions from this limited set of information it is necessary to 
include in the statistical analysis of numerical results 
also prior information on the structure of $G_H (\tau, T)$
as well as assumptions about the likelihood of a certain spectral
function $\sigma_H(\omega,T)$. It has been suggested to provide this
additional information through the application of the Maximum Entropy 
Method (MEM) \cite{Asa00,Nak99}, which has been applied successfully
to many other ill-conditioned problems in physics. In the context of QCD 
it has been
shown that the MEM analysis correctly describes spectral properties of 
Euclidean correlation functions at zero \cite{Asa00,Nak99} as well as 
finite \cite{Kar01} temperature.

We will discuss here results on thermal properties of light and
heavy quark mesonic correlation functions. We stress thermal properties
of the correlators, which independent of any statistical 
analysis provide evidence for changes in the spectral properties, and
in addition present results obtained from a MEM analysis of these 
correlation functions. 

\begin{figure*}[t]
\hspace*{-0.4cm}\epsfig{file=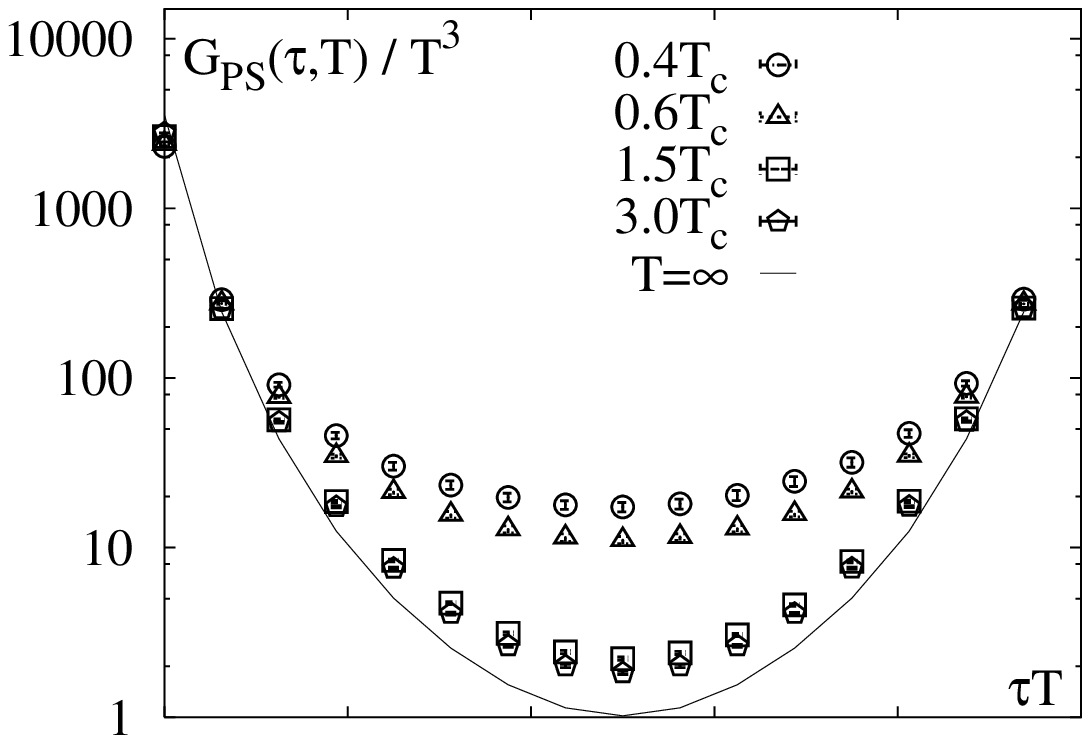,width=75mm}
\hspace*{-0.5cm}\epsfig{file=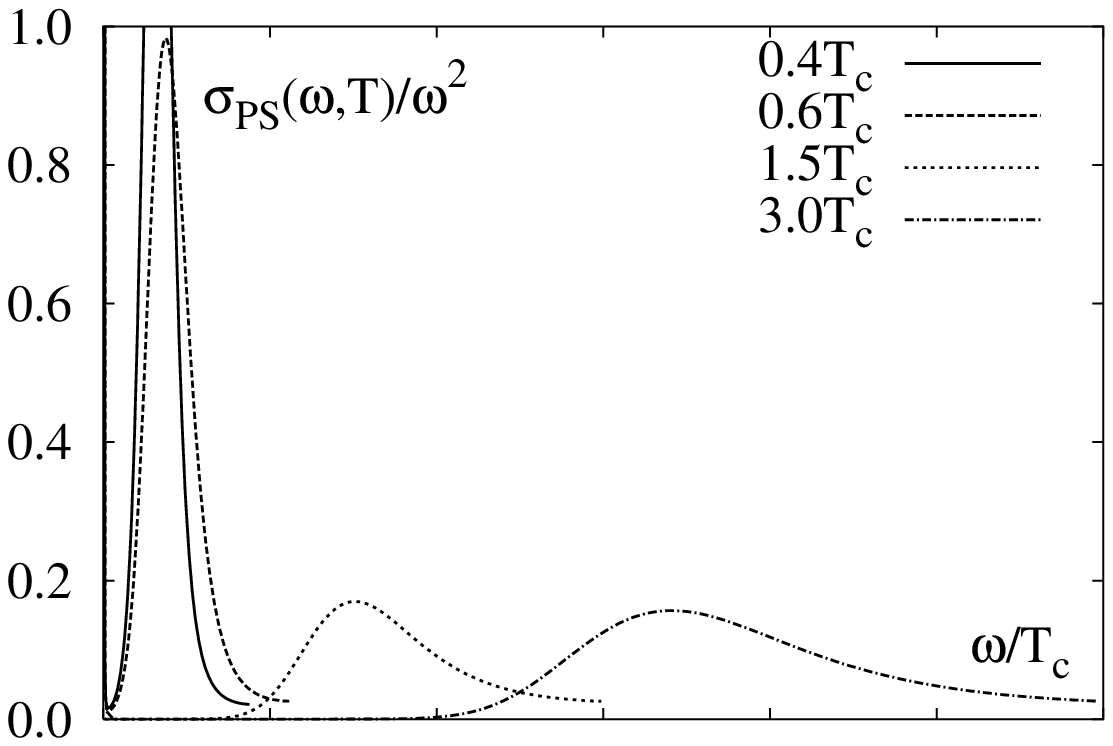,width=75mm}

\vskip -0.9truecm
\hspace*{-0.4cm}\epsfig{file=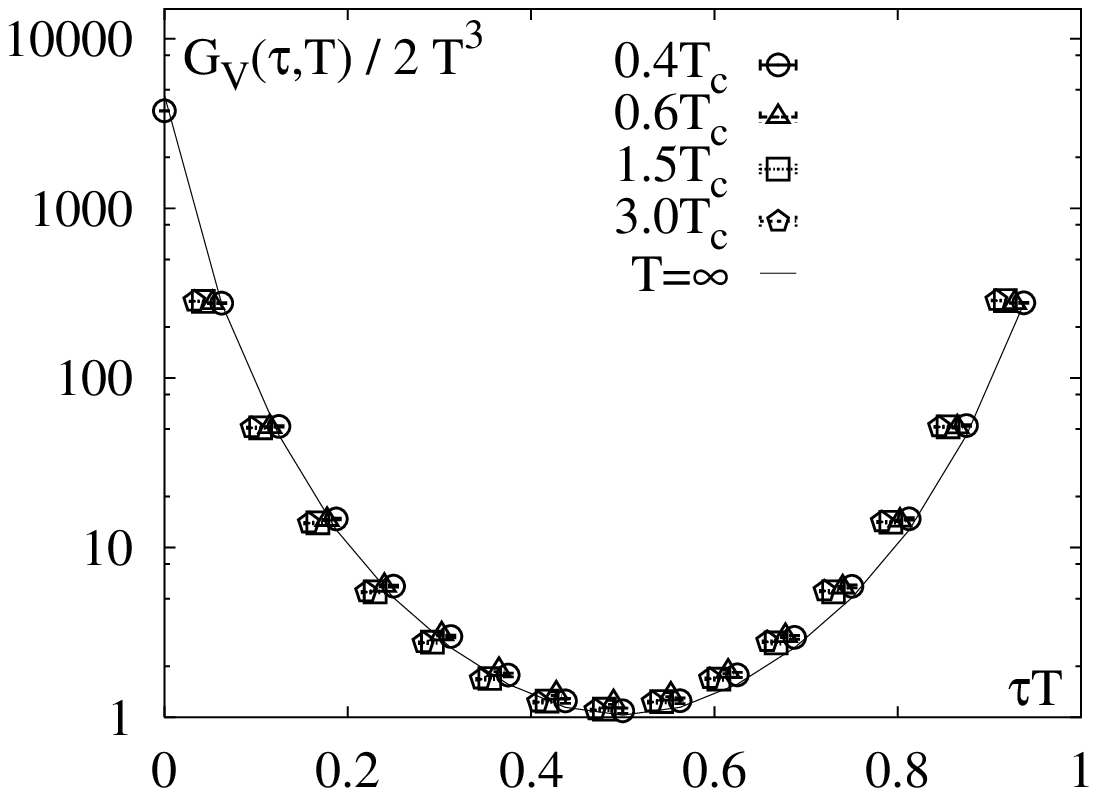,width=75mm}
\hspace*{-0.5cm}\epsfig{file=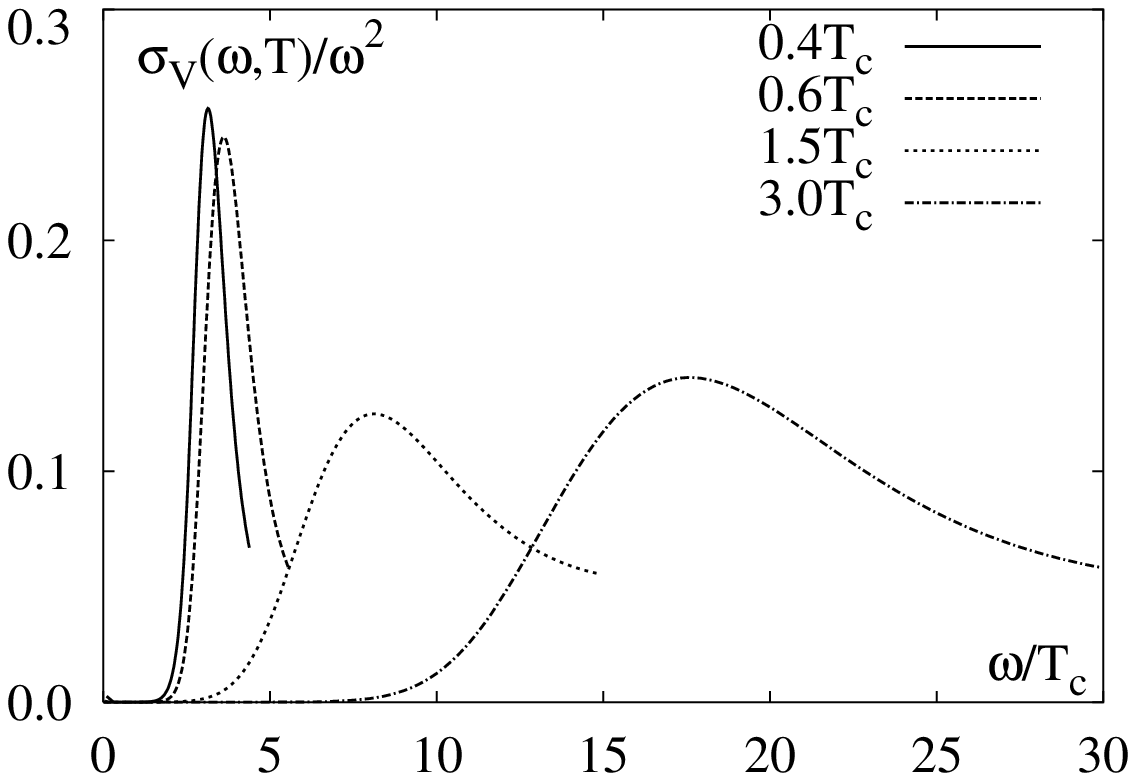,width=75mm}
\vskip -0.8truecm
\caption{Pseudo-scalar (up) and vector (down) correlation (left) and
spectral (right) functions. The solid line in the left hand figures
corresponds to the free ($T\rightarrow \infty$) mesonic correlation function 
calculated on 
lattices of same size. For better visibility we have displaced some
data for $G_V(\tau,T)$ and only show the low energy regime, $\omega/T <
10$, for $\sigma_H(\omega,T)$. 
Further details including an error analysis 
will be presented elsewhere.}
\label{lightmeson} 
\end{figure*}

\section{Euclidean correlation functions in the scalar and vector
channel} 

In Fig.~\ref{lightmeson} we show zero-momentum, pseudo-scalar (PS) and 
vector (V) meson correlation functions calculated at various temperatures 
on lattices of size $64^3\times 16$ in quenched QCD using
non-perturbatively improved Wilson fermions. Calculations below $T_c$
have been performed at non-vanishing values of the quark mass ($m_q$)
corresponding to a pion mass of about 500~MeV while above $T_c$
the calculations have been performed in the chiral limit. 
The drastic change that occurs in the structure of $G_H(\tau, T)$
when crossing the deconfinement transition temperature $T_c$ is 
clearly visible in the PS-channel. We note that this is not related
to the different $m_q$-values used below and above $T_c$. On the
contrary, decreasing $m_q$ further below $T_c$ would lead to 
even flatter correlation functions indicating a decrease in the 
PS-meson mass.
Indeed, a systematic analysis of the $m_q$-dependence shows that
the lowest excitation in the PS-channel remains massless in
the chiral limit for all temperatures below $T_c$. Above $T_c$ the 
scalar and pseudo-scalar correlation functions are degenerate,
reflecting approximate chiral ($U_A(1)$) symmetry restoration, and both 
approach the meson 
correlation function constructed from freely propagating,  massless quarks. 
However, in particular when we compare $G_{PS}(\tau, T)$ to the vector 
channel it is evident that it still deviates significantly
from this asymptotic form, which is expected to be reached only at 
infinite temperature. We also note that a free meson correlation function
constructed from massive quarks would be steeper than the solid curves
shown in the left hand part of Fig.~\ref{lightmeson}. The correlation
functions thus indicate that a strong correlation at low energies 
persists above $T_c$ for the propagating $q\bar{q}$-pair. 

The drastic change visible in $G_{PS}(\tau, T)$ also gets reflected in
the spectral functions reconstructed from this correlation function
using MEM. The pronounced $\delta$-function like peak present 
in the spectral functions below $T_c$ gets replaced by a broad
''resonance'' above $T_c$ which  moves to larger energies with increasing $T$.
In fact, the spectral functions almost coincide above $T_c$ when plotted 
as function of $\omega /T$ (rather than $\omega /T_c$ as shown in
Fig.~\ref{lightmeson}) which reflects the weak temperature dependence
of $G_{PS}(\tau, T)$ above $T_c$. 

The structure of the spectral functions obtained in the vector channel
is similar to those in the pseudo-scalar channel, although the
vector correlation function is much closer to the free vector
correlation function at all temperatures. In the spectral function this
is reflected in the reduced strength of the vector meson pole
contribution below $T_c$ (note the different scale in the right
hand figures) and a less pronounced ''resonance'' like structure
above $T_c$. 

\section{Thermal dilepton rate}

The vector spectral function shown in Fig.~\ref{lightmeson} is directly
related to the thermal cross section for the production of dilepton
pairs at vanishing momentum (see Fig.~2).  

\vspace{1.5cm}
\noindent
$\displaystyle{{{\rm d} W \over {\rm d}\omega {\rm d}^3p} =
{5 \alpha^2 \over 27 \pi^2} {1\over \omega^2 ({\rm e}^{\omega/T} - 1)}
\sigma_V(\omega,\vec{p},T)} $
\vspace{-1.5cm}
\begin{figure*}[h]
\vskip -1.8truecm
\begin{center}
\hfill\epsfig{file=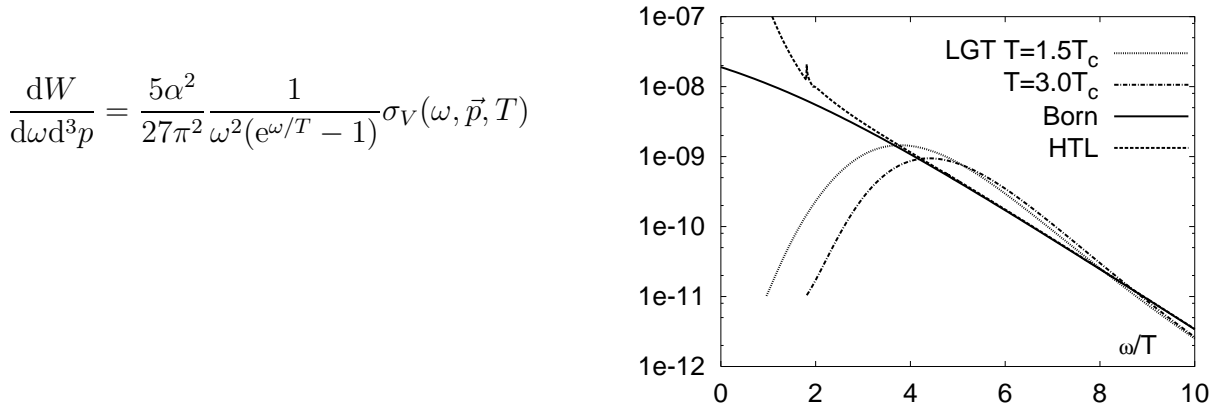,width=84mm}
\end{center}
\vskip -1.2truecm
\caption{Thermal dilepton cross section determined from the vector
spectral function calculated on the lattice as shown in Fig.~1. For 
comparison we also give the leading order perturbative result (Born)
and the result of the HTL-resummed perturbative calculation
\cite{Braaten} with a thermal mass, $m_{HTL}/T\;=\;g(T)/\sqrt{6}$,
chosen as unity.}
\label{rate}
\end{figure*}

This thermal dilepton rate is shown in Fig.~\ref{rate}. The
``resonance'' like enhancement is visible in this figure as an
enhancement of the rate over the perturbative tree level (Born) rate for    
energies $\omega / T \in [4,8]$. Discrepancies with hard thermal loop
calculations \cite{Braaten} show up at smaller energies where lattice 
results for
the spectral functions rapidly drop while the HTL result diverges in 
the infrared limit. It is to be expected that perturbative as well
as lattice calculations do not yet describe this infrared regime
correctly and will improve in the future. The fact, that the lattice 
calculations of $G_V(1/2T, T)$ do stay close to
the free result, $G_V^{\rm free}(1/2T, T) \equiv 2$, also above $T_c$ 
does, however, put already now a 
stringent constraint on the properties of $\sigma_V(\omega, T)$. In 
particular, the value of $G_V(\tau, T)$ at $\tau T = 1/2$ is related to
a simple, exponentially damped integral over the spectral function,

\begin{equation}
{G_V (1/2T, T) \over T^3} \equiv 
\int_0^{\infty} \hspace*{-0.2cm}{\rm d} \omega
\;
{\sigma_V(\omega,T)\; \over {\rm sinh} (\omega/2T)}
= \cases{
2.23 \pm 0.05 &, $T/T_c = 1.5$ \cr
2.21 \pm 0.05 &, $T/T_c = 3$}
\quad .
\label{cor2T}
\end{equation}
This relation demands that the spectral function vanishes 
in the limit $\omega \rightarrow  0$. In fact, in order to get
non-vanishing, finite transport coefficients in the QGP the spectral
function should be proportional to $\omega$ in this limit \cite{Aarts}.

\section{Heavy quark bound states at high temperature}

As the properties of mesons constructed from light quarks are closely 
related to chiral properties of QCD it is expected that these states
are strongly influenced by the chiral phase transition to the 
QGP. The situation, however, is different for heavy quark
bound states, which generally are expected to be sensitive to the
deconfining aspect of the QCD phase transition. Whether 
a heavy quark bound state survives the QCD phase transition 
or not strongly depends on the efficient screening of the interaction
among quarks and anti-quarks in the QGP. Potential model calculations
indicate that some $c\bar{c}$ bound states, e.g. the $J/\psi$,
could survive at temperatures close to $T_c$ while orbitally excited
states like the $\chi_c$ most likely get dissolved at $T_c$ \cite{Satz}.   
A more direct analysis of the fate of heavy quark bound states again
can come from the analysis of thermal meson correlation functions
constructed with heavy quarks. Like in the light quark sector already
the temperature dependence of the correlation functions themselves is
instructive. 
The pseudo-scalar ($\eta_c)$ and vector ($J/\psi$) correlation 
functions show almost no temperature dependence across the phase 
transition while the scalar ($\chi_{c,0}$) and axial-vector 
($\chi_{c,1}$) change significantly. This also is 
reflected in the spectral functions extracted from these correlators
\cite{Dat02}. They suggest that the scalar and 
axial-vector states dissolve at $T_c$ while the pseudo-scalar
and vector states still seem to be bound at $1.5\; T_c$.

\end{document}